\documentclass[11pt]{article}
\usepackage{amsmath,amssymb,amsfonts,bbm}

\newcommand{\be}{\begin{equation}}
\newcommand{\ee}{\end{equation}}
\newcommand{\bea}{\begin{eqnarray}}
\newcommand{\eea}{\end{eqnarray}}
\newcommand{\ba}{\begin{array}}
\newcommand{\ea}{\end{array}}

\newcommand{\htwo}{h_{2,1}}
\newcommand{\M}{\mathcal{M}}
\newcommand{\N}{\mathcal{N}}

\newcommand{\K}{\mathcal{K}}

\long\def\symbolfootnote[#1]#2{\begingroup%
\def\thefootnote{\fnsymbol{footnote}}\footnote[#1]{#2}\endgroup}

\begin{document}

\thispagestyle{empty}\vspace{40pt}

\hfill{}

\vspace{128pt}

\begin{center}
    \textbf{\Large Zero-branes and the symplectic hypermultiplets}\\
    \vspace{40pt}

    Moataz H. Emam\symbolfootnote[1]{\tt moataz.emam@cortland.edu}

    \vspace{12pt}   \textit{Department of Physics}\\
                    \textit{SUNY College at Cortland}\\
                    \textit{Cortland, New York 13045, USA}\\
\end{center}

\vspace{40pt}

\begin{abstract}

We study the scalar fields of the five-dimensional $\N=2$ hypermultiplets using the method of symplectic covariance developed in previous work. For static spherically symmetric backgrounds, we show that exactly two possibilities exist. One of them is a Bogomol’nyi-Prasad-Sommerfeld (BPS) zero-brane carrying charge under the hypermultiplets. We find an explicitly symplectic solution of the fields in this background and derive the conditions required for a full spacetime understanding.

\end{abstract}

\newpage



\vspace{15pt}

\pagebreak

\section{Introduction}

The study of $\N=2$ supergravity theories in four as well as five dimensions is a popular enterprise in the literature. It is generally motivated by these theories' possible roles in understanding the string-theoretic origins of entropy, duality symmetries, the AdS/CFT correspondence, as well as the structure of the underlying special holonomy manifolds; in our case a Calabi-Yau (CY) 3-fold. Classifications of solutions and \emph{almost} solutions (i.e. constructions with constraints) exist in abundance, (for example \cite{Butter:2012xg,Klemm:2009uw,Mohaupt:2008zz,Cacciatori:2008ek,Cortes:2003zd,Isozumi:2003uh,Cacciatori:2003kv,Andrianopoli:2011zj} and references within). It should however be noted that the vast majority of such studies focus on the vector and/or tensor multiplet regimes. Comparatively, little work is being done on the hypermultiplets sector. This is due, in part, to the mathematical complexity involved, since the hypermultiplets generally parameterize quaternionic manifolds \cite{de Wit:2001dj}. However, it was pointed out some years ago that due to the so-called \textit{c}-map, the hypermultiplets in $D=5$ for instance can be related to the much better understood $D=4$ vector multiplets, and that the methods of special geometry, developed for the latter, can be applied to the former \cite{Gutperle:2000ve}. Based on this observation, some hypermultiplet constructions in instanton and certain two-brane backgrounds were found and studied (last reference and \cite{Emam:2005bh,Emam:2006sr}). Despite this, explicit calculations remain tedious even in the relatively simpler language of special geometry (as compared to the original quaternionic language). We argued in \cite{Emam:2009xj} that the well-known symplectic structure of quaternionic and special  K\"{a}hler manifolds can be used to construct hypermultiplet `solutions' based on covariance in symplectic space. These are full solutions only in the symplectic sense, written in terms of symplectic basis vectors and invariants. As far as being spacetime solutions however, they are only partial, or almost-solutions, in the sense that they depend on the unknown explicit form of the underlying Calabi-Yau. In fact, we have also argued that the reverse can be true: constraint equations derived from these `solutions' may eventually lead to an understanding of the submanifold itself. The symplectic structure of special geometry is, of course, a known property and is well understood (e.g. \cite{deWit:1995jd}). Our contribution was simply to use it to recast the theory into a symplectic form that greatly reduces the work needed as well as provide a technique by which to construct solutions. Previously, we have only shown the application of this to results that were already known in the literature, found by the methods of special geometry \cite{Gutperle:2000ve,Emam:2005bh}. In the current work, we apply the symplectic method to show that, under certain constraints of maximal symmetry (spherically symmetric backgrounds etc), there exist $D=5$ BPS zero-brane solutions coupled to the most general form of the $\N=2$ hypermultiplets. We find some explicit spacetime solutions to the hyperscalars and derive constraints on the complex structure moduli of the underlying Calabi-Yau that may be used in future work for a deeper understanding.

The paper is organized as follows: In section (\ref{theory}) we review the form of ungauged $\N=2$ supergravity theory in five dimensions using the symplectic formulation. In section (\ref{Brane Analysis}) we analyze the Einstein equation and the BPS conditions to show that for a general spherically symmetric static $p$-brane background only two possibilities of $p$ exist. Finally in section (\ref{Fields}) we look at the case $p=0$ in some detail.

\section{$D=5$ $\N=2$ supergravity with hypermultiplets} \label{theory}

The dimensional reduction of $D=11$ supergravity theory over a Calabi-Yau 3-fold $\M$ with nontrivial complex structure moduli yields an $\N=2$ supergravity theory in $D=5$ with a set of scalar fields and their supersymmetric partners all together known as the \emph{hypermultiplets} (see \cite{Emam:2010kt} for a review and additional references). It should be noted that the other matter sector in the theory; the vector multiplets, trivially decouples from the hypermultiplets and can be simply set to zero, as we do here. The hypermultiplets are partially comprised of the \emph{universal hypermultiplet} $\left(a, \sigma, \zeta^0, \tilde \zeta_0\right)$; so called because it appears irrespective of the detailed structure of the Calabi-Yau. The field $a$ is known as the universal axion, and is magnetically dual to a three-form gauge field and the dilaton $\sigma$ is proportional to the natural logarithm of the volume of $\M$. The rest of the hypermultiplets are $\left(z^i, z^{\bar i}, \zeta^i, \tilde \zeta_i: i=1,\ldots, \htwo\right)$, where the $z$'s are identified with the complex structure moduli of $\M$, and $\htwo$ is the Hodge number determining the dimensions of the manifold of the Calabi-Yau's complex structure moduli, $\M_C$. The `bar' over an index denotes complex conjugation. The fields $\left(\zeta^I, \tilde\zeta_I: I=0,\ldots,\htwo\right)$ are known as the axions and arise as a result of the $D=11$ Chern-Simons term. The supersymmetric partners known as the hyperini complete the hypermultiplets.

The theory has a very rich structure that arises from the intricate topology of $\M$. Of particular interest to us is its symplectic covariance. Particularly, the axions $\left(\zeta^I, \tilde\zeta_I\right)$ can be defined as components of the symplectic vector
\be\label{DefOfSympVect}
   \left| \Xi  \right\rangle  = \left( {\begin{array}{*{20}c}
   {\,\,\,\,\,\zeta ^I }  \\
   -{\tilde \zeta _I }  \\
    \end{array}} \right),
\ee
such that the symplectic scalar product is defined by, for example,
\be
    \left\langle {{\Xi }}
 \mathrel{\left | {\vphantom {{\Xi } d\Xi }}
 \right. \kern-\nulldelimiterspace}
 {d\Xi } \right\rangle   = \zeta^I d\tilde \zeta_I  - \tilde \zeta_I
 d\zeta^I,\label{DefOfSympScalarProduct}
\ee
where $d$ is the spacetime exterior derivative $\left(d=dx^M\partial_M:M=0,\ldots,4\right)$. A `rotation' in symplectic space is defined by the matrix element
\bea
 \left\langle {\partial _M \Xi } \right|{\bf\Lambda} \left| {\partial ^M \Xi } \right\rangle \star \mathbf{1} &=& \left\langle {d\Xi } \right|\mathop {\bf\Lambda} \limits_ \wedge  \left| {\star d\Xi } \right\rangle  \nonumber\\
  &=& 2\left\langle {{d\Xi }}
 \mathrel{\left | {\vphantom {{d\Xi } V}}
 \right. \kern-\nulldelimiterspace}
 {V} \right\rangle \mathop {}\limits_ \wedge  \left\langle {{\bar V}}
 \mathrel{\left | {\vphantom {{\bar V} {\star d\Xi }}}
 \right. \kern-\nulldelimiterspace}
 {{\star d\Xi }} \right\rangle  + 2G^{i\bar j} \left\langle {{d\Xi }}
 \mathrel{\left | {\vphantom {{d\Xi } {U_{\bar j} }}}
 \right. \kern-\nulldelimiterspace}
 {{U_{\bar j} }} \right\rangle \mathop {}\limits_ \wedge  \left\langle {{U_i }}
 \mathrel{\left | {\vphantom {{U_i } {\star d\Xi }}}
 \right. \kern-\nulldelimiterspace}
 {{\star d\Xi }} \right\rangle  - i\left\langle {d\Xi } \right.\mathop |\limits_ \wedge  \left. {\star d\Xi } \right\rangle,\label{DefOfRotInSympSpace}
\eea
where $\star$ is the $D=5$ Hodge duality operator, and $G_{i\bar j}$ is a special K\"{a}hler metric on $\M_C$. The symplectic basis vectors $\left| V \right\rangle $, $\left| {U_i } \right\rangle $ and their complex conjugates are defined by
\be
    \left| V \right\rangle  = e^{\frac{\K}{2}} \left( {\begin{array}{*{20}c}
   {Z^I }  \\
   {F_I }  \\
    \end{array}} \right),\,\,\,\,\,\,\,\,\,\,\,\,\,\,\,\left| {\bar V} \right\rangle  = e^{\frac{\K}{2}} \left( {\begin{array}{*{20}c}
   {\bar Z^I }  \\
   {\bar F_I }  \\
    \end{array}} \right)\label{DefOfVAndVBar}
\ee
where $\K$ is the K\"{a}hler potential on $\M_C$, $\left( {Z,F} \right)$ are the periods of the Calabi-Yau's holomorphic volume form, and
\bea
    \left| {U_i } \right\rangle  &=& \left| \nabla _i V
    \right\rangle=\left|\left[ {\partial _i  + \frac{1}{2}\left( {\partial _i \K} \right)} \right] V \right\rangle \nonumber\\
    \left| {U_{\bar i} } \right\rangle  &=& \left|\nabla _{\bar i}  {\bar V} \right\rangle=\left|\left[ {\partial _{\bar i}  + \frac{1}{2}\left( {\partial _{\bar i} \K} \right)} \right] {\bar V}
    \right\rangle\label{DefOfUAndUBar}
\eea
where the derivatives are with respect to the moduli $\left(z^i, z^{\bar i}\right)$. These vectors satisfy the following conditions:
\bea
    \left\langle {{\bar V}}
     \mathrel{\left | {\vphantom {{\bar V} V}}
     \right. \kern-\nulldelimiterspace}
     {V} \right\rangle   &=& i\nonumber\\
    \left|\nabla _i  {\bar V} \right\rangle  &=& \left|\nabla _{\bar i}  V \right\rangle =0\nonumber\\
    \left\langle {{U_i }}
    \mathrel{\left | {\vphantom {{U_i } {U_j }}}
    \right. \kern-\nulldelimiterspace}
    {{U_j }} \right\rangle  &=& \left\langle {{U_{\bar i} }}
    \mathrel{\left | {\vphantom {{U_{\bar i} } {U_{\bar j} }}}
    \right. \kern-\nulldelimiterspace}
    {{U_{\bar j} }} \right\rangle    =0\nonumber\\
    \left\langle {\bar V}
    \mathrel{\left | {\vphantom {\bar V {U_i }}}
    \right. \kern-\nulldelimiterspace}
    {{U_i }} \right\rangle  &=& \left\langle {V}
    \mathrel{\left | {\vphantom {V {U_{\bar i} }}}
    \right. \kern-\nulldelimiterspace}
    {{U_{\bar i} }} \right\rangle  = \left\langle { V}
    \mathrel{\left | {\vphantom { V {U_i }}}
    \right. \kern-\nulldelimiterspace}
    {{U_i }} \right\rangle=\left\langle {\bar V}
    \mathrel{\left | {\vphantom {\bar V {U_{\bar i} }}}
    \right. \kern-\nulldelimiterspace}
    {{U_{\bar i} }} \right\rangle= 0,\nonumber\\
    \left|\nabla _{\bar j}  {U_i } \right\rangle  &=& G_{i\bar j} \left| V \right\rangle ,\quad \quad \left|\nabla _i  {U_{\bar j} } \right\rangle  = G_{i\bar j} \left| {\bar V}
    \right\rangle,\nonumber\\
    G_{i\bar j}&=& \left( {\partial _i \partial _{\bar j} \K} \right)=- i    \left\langle {{U_i }}
    \mathrel{\left | {\vphantom {{U_i } {U_{\bar j} }}}
    \right. \kern-\nulldelimiterspace}
    {{U_{\bar j} }} \right\rangle.
\eea

The origin of these identities lies in special K\"{a}hler geometry. In our previous work \cite{Emam:2009xj}, we derived the following useful formulae:
\bea
    dG_{i\bar j}  &=& G_{k\bar j} \Gamma _{ri}^k dz^r  + G_{i\bar k} \Gamma _{\bar r\bar j}^{\bar k} dz^{\bar r}  \nonumber\\
    dG^{i\bar j}  &=&  - G^{p\bar j} \Gamma _{rp}^i dz^r  - G^{i\bar p} \Gamma _{\bar r\bar p}^{\bar j} dz^{\bar r}  \nonumber\\
    \left| {dV} \right\rangle  &=& dz^i \left| {U_i } \right\rangle  - i\mathcal{P}\left| V \right\rangle \nonumber \\
    \left| {d\bar V} \right\rangle  &=& dz^{\bar i} \left| {U_{\bar i} } \right\rangle  + i\mathcal{P}\left| {\bar V} \right\rangle \nonumber
\eea
\bea
    \left| {dU_i } \right\rangle  &=& G_{i\bar j} dz^{\bar j} \left| V \right\rangle  + \Gamma _{ik}^r dz^k \left| {U_r } \right\rangle+G^{j\bar l} C_{ijk} dz^k \left| {U_{\bar l} } \right\rangle - i\mathcal{P}\left| {U_i } \right\rangle \nonumber \\
    \left| {dU_{\bar i} } \right\rangle  &=& G_{j\bar i} dz^j \left| {\bar V} \right\rangle + \Gamma _{\bar i\bar k}^{\bar r} dz^{\bar k} \left| {U_{\bar r} } \right\rangle + G^{l\bar j} C_{\bar i\bar j\bar k} dz^{\bar k} \left| {U_l } \right\rangle + i\mathcal{P}\left| {U_{\bar i} } \right\rangle\nonumber
\eea
\bea
     {\bf \Lambda } &=& 2\left| V \right\rangle \left\langle {\bar V} \right| + 2G^{i\bar j} \left| {U_{\bar j} } \right\rangle \left\langle {U_i } \right|
    -i\nonumber\\
    {\bf \Lambda }^{-1} &=& -2\left| V \right\rangle \left\langle {\bar V} \right| - 2G^{i\bar j} \left| {U_{\bar j} } \right\rangle \left\langle {U_i } \right|
    +i\nonumber\\
        \partial_i {\bf \Lambda } &=& 2\left| {U_i } \right\rangle \left\langle {\bar V} \right|+2\left| {\bar V} \right\rangle \left\langle {U_i } \right| + 2G^{j\bar r} G^{k\bar p} C_{ijk} \left| {U_{\bar r} } \right\rangle \left\langle {U_{\bar p} } \right|
\eea
where
\be
 \mathcal{P} =
\mathfrak{Im} \left[ {\left( {\partial_i  \K} \right)dz^i} \right].
\ee

The quantities $C_{ijk}$ are the components of the totally symmetric tensor that appears in the curvature tensor of $\M_C$. In this language, the bosonic part of the action is:
\bea
    S_5  &=& \int\limits_5 {\left[ {R\star \mathbf{1} - \frac{1}{2}d\sigma \wedge\star d\sigma  - G_{i\bar j} dz^i \wedge\star dz^{\bar j} } \right.}  + e^\sigma   \left\langle {d\Xi } \right|\mathop {\bf\Lambda} \limits_ \wedge  \left| {\star d\Xi } \right\rangle\nonumber\\
    & &\left. {\quad\quad\quad\quad\quad\quad\quad\quad\quad\quad\quad\quad\quad - \frac{1}{2} e^{2\sigma } \left[ {da + \left\langle {\Xi } \mathrel{\left | {\vphantom {\Xi  {d\Xi }}} \right. \kern-\nulldelimiterspace} {{d\Xi }}    \right\rangle} \right] \wedge \star\left[ {da + \left\langle {\Xi } \mathrel{\left | {\vphantom {\Xi  {d\Xi }}} \right. \kern-\nulldelimiterspace} {{d\Xi }}    \right\rangle} \right] } \right].\label{action}
\eea

The variation of the action yields the following field equations for $\sigma$, $\left(z^i,z^{\bar i}\right)$, $\left| \Xi  \right\rangle$ and $a$ respectively:
\bea
    \left( {\Delta \sigma } \right)\star \mathbf{1} + e^\sigma   \left\langle {d\Xi } \right|\mathop {\bf\Lambda} \limits_ \wedge  \left| {\star d\Xi } \right\rangle -   e^{2\sigma }\left[ {da + \left\langle {\Xi } \mathrel{\left | {\vphantom {\Xi  {d\Xi }}} \right. \kern-\nulldelimiterspace} {{d\Xi }}    \right\rangle} \right]\wedge\star\left[ {da + \left\langle {\Xi } \mathrel{\left | {\vphantom {\Xi  {d\Xi }}} \right. \kern-\nulldelimiterspace} {{d\Xi }}    \right\rangle} \right] &=& 0\label{DilatonEOM}\\
    \left( {\Delta z^i } \right)\star \mathbf{1} + \Gamma _{jk}^i dz^j  \wedge \star dz^k  + \frac{1}{2}e^\sigma  G^{i\bar j}  {\partial _{\bar j} \left\langle {d\Xi } \right|\mathop {\bf\Lambda} \limits_ \wedge  \left| {\star d\Xi } \right\rangle} &=& 0 \nonumber\\
    \left( {\Delta z^{\bar i} } \right)\star \mathbf{1} + \Gamma _{\bar j\bar k}^{\bar i} dz^{\bar j}  \wedge \star dz^{\bar k}  + \frac{1}{2}e^\sigma  G^{\bar ij}  {\partial _j \left\langle {d\Xi } \right|\mathop {\bf\Lambda} \limits_ \wedge  \left| {\star d\Xi } \right\rangle}  &=& 0\label{ZZBarEOM} \\
    d^{\dag} \left\{ {e^\sigma  \left| {{\bf\Lambda} d\Xi } \right\rangle  - e^{2\sigma } \left[ {da + \left\langle {\Xi }
    \mathrel{\left | {\vphantom {\Xi  {d\Xi }}}\right. \kern-\nulldelimiterspace} {{d\Xi }} \right\rangle } \right]\left| \Xi  \right\rangle } \right\} &=& 0\label{AxionsEOM}\\
    d^{\dag} \left[ {e^{2\sigma } da + e^{2\sigma } \left\langle {\Xi } \mathrel{\left | {\vphantom {\Xi  {d\Xi }}} \right. \kern-\nulldelimiterspace} {{d\Xi }}    \right\rangle} \right] &=&    0\label{aEOM}
\eea
where $d^\dagger$ is the $D=5$ adjoint exterior derivative, $\Delta$ is the Laplace-de Rahm operator and $\Gamma _{jk}^i$ is a connection on $\M_C$. The full action is symmetric under the following SUSY transformations:
\bea
 \delta _\epsilon  \psi ^1  &=& D \epsilon _1  + \frac{1}{4}\left\{ {i {e^{\sigma } \left[ {da + \left\langle {\Xi }
 \mathrel{\left | {\vphantom {\Xi  {d\Xi }}}
 \right. \kern-\nulldelimiterspace} {{d\Xi }} \right\rangle } \right]}- Y} \right\}\epsilon _1  - e^{\frac{\sigma }{2}} \left\langle {{\bar V}}
 \mathrel{\left | {\vphantom {{\bar V} {d\Xi }}} \right. \kern-\nulldelimiterspace} {{d\Xi }} \right\rangle\epsilon _2  \nonumber\\
 \delta _\epsilon  \psi ^2  &=& D \epsilon _2  - \frac{1}{4}\left\{ {i {e^{\sigma } \left[ {da + \left\langle {\Xi }
 \mathrel{\left | {\vphantom {\Xi  {d\Xi }}} \right. \kern-\nulldelimiterspace}
 {{d\Xi }} \right\rangle } \right]}- Y} \right\}\epsilon _2  + e^{\frac{\sigma }{2}} \left\langle {V}
 \mathrel{\left | {\vphantom {V {d\Xi }}} \right. \kern-\nulldelimiterspace} {{d\Xi }} \right\rangle \epsilon _1  \label{SUSYGraviton}
\eea
\bea
  \delta _\epsilon  \xi _1^0  &=& e^{\frac{\sigma }{2}} \left\langle {V}
    \mathrel{\left | {\vphantom {V {\partial _M  \Xi }}} \right. \kern-\nulldelimiterspace} {{\partial _M  \Xi }} \right\rangle  \Gamma ^M  \epsilon _1  - \left\{ {\frac{1}{2}\left( {\partial _M  \sigma } \right) - \frac{i}{2} e^{\sigma } \left[ {\left(\partial _M a\right) + \left\langle {\Xi }
    \mathrel{\left | {\vphantom {\Xi  {\partial _M \Xi }}} \right. \kern-\nulldelimiterspace}
    {{\partial _M \Xi }} \right\rangle } \right]} \right\}\Gamma ^M  \epsilon _2  \nonumber\\
     \delta _\epsilon  \xi _2^0  &=& e^{\frac{\sigma }{2}} \left\langle {{\bar V}}
    \mathrel{\left | {\vphantom {{\bar V} {\partial _M  \Xi }}} \right. \kern-\nulldelimiterspace} {{\partial _M  \Xi }} \right\rangle \Gamma ^M  \epsilon _2  + \left\{ {\frac{1}{2}\left( {\partial _M  \sigma } \right) + \frac{i}{2} e^{\sigma } \left[ {\left(\partial _M a\right) + \left\langle {\Xi }
    \mathrel{\left | {\vphantom {\Xi  {\partial _M \Xi }}} \right. \kern-\nulldelimiterspace}
    {{\partial _M \Xi }} \right\rangle } \right]} \right\}\Gamma ^M  \epsilon
     _1\label{SUSYHyperon1}
\eea
\bea
     \delta _\epsilon  \xi _1^{\hat i}  &=& e^{\frac{\sigma }{2}} e^{\hat ij} \left\langle {{U_j }}
    \mathrel{\left | {\vphantom {{U_j } {\partial _M  \Xi }}} \right. \kern-\nulldelimiterspace} {{\partial _M  \Xi }} \right\rangle \Gamma ^M  \epsilon _1  - e_{\,\,\,\bar j}^{\hat i} \left( {\partial _M  z^{\bar j} } \right)\Gamma ^M  \epsilon _2  \nonumber\\
     \delta _\epsilon  \xi _2^{\hat i}  &=& e^{\frac{\sigma }{2}} e^{\hat i\bar j} \left\langle {{U_{\bar j} }}
    \mathrel{\left | {\vphantom {{U_{\bar j} } {\partial _M  \Xi }}} \right. \kern-\nulldelimiterspace} {{\partial _M  \Xi }} \right\rangle \Gamma ^M  \epsilon _2  + e_{\,\,\,j}^{\hat i} \left( {\partial _M  z^j } \right)\Gamma ^M  \epsilon    _1,\label{SUSYHyperon2}
\eea
where $\left(\psi ^1, \psi ^2\right)$ are the two gravitini and $\left(\xi _1^I, \xi _2^I\right)$ are the hyperini. The quantity $Y$ is defined by:
\begin{equation}
    Y   = \frac{{\bar Z^I N_{IJ}  {d  Z^J }  -
    Z^I N_{IJ}  {d  \bar Z^J } }}{{\bar Z^I N_{IJ} Z^J
    }},\label{DefOfY}
\end{equation}
where $N_{IJ}  = \mathfrak{Im} \left({\partial_IF_J } \right)$. The $e$'s are the beins of the special K\"{a}hler metric $G_{i\bar j}$, the $\epsilon$'s are the five-dimensional $\N=2$ SUSY spinors and the $\Gamma$'s are the usual Dirac matrices. Finally, the covariant derivative $D$ is given by
\be
    D=dx^M\left( \partial _M   + \frac{1}{4}\omega _M^{\,\,\,\,\hat M\hat N} \Gamma _{\hat M\hat
    N}\right)\label{DefOfCovDerivative}
\ee
as usual, where the $\omega$'s are the spin connections and the hatted indices are frame indices in a flat tangent space.

\section{Brane analysis}\label{Brane Analysis}

The most general spherically symmetric $p$-branes in $D=5$ can be represented by the following (Poincar\'{e})$_{p+1} \times SO\left(4-p\right)$ metric:
\be
    ds^2  = e^{2C\sigma } \eta _{ab} dx^a dx^b  + e^{2B\sigma } \delta _{\mu \nu } dx^\mu  dx^\nu,\label{GeneralMetric}
\ee
where $B$ and $C$ are constants, the directions $a,b=0,1,\ldots,p$ define the brane's world-volume while $\mu,\nu=(p+1),\ldots,4$ are those transverse to the brane. The dilaton is assumed purely radial in the $\mu,\nu$ directions. It turns out that the constant $C$ is constrained to vanish by both the Einstein equations and the SUSY condition\footnote{This is no surprise, since $\delta \psi=0$ automatically satisfies $G_{MN}= T_{MN}$ \cite{Kaya:1999mm}.} $\delta \psi=0$. Looking ahead, this can be easily seen by considering the $b$ components of $\delta \psi=0$ and noting that all terms but one vanish because of the fields' independence of $x^b$:
\bea
    \partial_b\epsilon _1   + \frac{C}{2}\left( {\partial _\nu  \sigma} \right){\Gamma _b } ^\nu\epsilon _1 + \frac{1}{4}\left\{ {i {e^{\sigma } \left[ {\partial_b a + \left\langle {\Xi } \mathrel{\left | {\vphantom {\Xi  {\partial_b\Xi }}} \right. \kern-\nulldelimiterspace} {{\partial_b\Xi }} \right\rangle } \right]}- Y_b} \right\}\epsilon _1  - e^{\frac{\sigma }{2}} \left\langle {{\bar V}} \mathrel{\left | {\vphantom {{\bar V} {\partial_b\Xi }}} \right. \kern-\nulldelimiterspace} {{\partial_b\Xi }} \right\rangle\epsilon _2 &=&0\nonumber\\
    {\rm leading} \,\,{\rm to}\quad\quad\quad \frac{C}{2}\left( {\partial _\nu  \sigma} \right){\Gamma _b } ^\nu \epsilon _1&=&0.
    \label{WhyC=0}
\eea

We then set $C=0$ from the start and are left only with the task of specifying $B$ and the allowed values of $p$. Based on this metric, the Ricci tensor breaks up into
\bea
 R_{ab}  &=& 0 \nonumber\\
 R_{\mu \nu }  &=&  - B\left( {2 - p} \right)\left( {\partial _\mu  \partial _\nu  \sigma } \right) - B\delta _{\mu \nu } \delta ^{\alpha \beta } \left( {\partial _\alpha  \partial _\beta  \sigma } \right) \nonumber\\
  & & \, \, \, B^2 \left( {2 - p} \right)\left( {\partial _\mu  \sigma } \right)\left( {\partial _\nu  \sigma } \right) - B^2 \left( {2 - p} \right)g _{\mu \nu }  \left( {\partial _\alpha  \sigma } \right)\left( {\partial ^\alpha  \sigma } \right)
\eea\label{RicciTensor}
leading to the Einstein tensor
\bea
 G_{ab}  &=& B\left( {3 - p} \right)\eta _{ab} g ^{\mu \nu } \left( {\partial _\mu  \partial _\nu  \sigma } \right) + \frac{1}{2}B^2 \left( {2 - p} \right)\left( {3 - p} \right)\eta _{ab}  \left( {\partial _\alpha  \sigma } \right)\left( {\partial ^\alpha  \sigma } \right) \nonumber\\
 G_{\mu \nu }  &=&  - B\left( {2 - p} \right)\left( {\partial _\mu  \partial _\nu  \sigma } \right) + B\left( {2 - p} \right)\delta _{\mu \nu } \delta ^{\alpha \beta } \left( {\partial _\alpha  \partial _\beta  \sigma } \right) \nonumber\\
& &  + \frac{1}{2}B^2 \left( {1 - p} \right)\left( {2 - p} \right)g _{\mu \nu }  \left( {\partial _\alpha  \sigma } \right)\left( {\partial ^\alpha  \sigma } \right) + B^2 \left( {2 - p} \right)\left( {\partial _\mu  \sigma } \right)\left( {\partial _\nu  \sigma } \right).\label{EinsteinTensor}
\eea

Variation of the matter part of the action with respect to the metric yields the stress tensor
\bea
 T_{ab}  &=&   \frac{1}{4}\eta _{ab} \left( {\partial _\alpha  \sigma } \right)\left( {\partial ^\alpha  \sigma } \right) + \frac{1}{2}\eta _{ab} G_{i\bar j} \left( {\partial _\alpha  z^i } \right)\left( {\partial ^\alpha  z^{\bar j} } \right) \nonumber\\
 & & -\frac{1}{2}\eta _{ab} e^\sigma  \left\langle {\partial _\alpha \Xi } \right|{\bf\Lambda} \left| {\partial ^\alpha \Xi } \right\rangle + \frac{1}{4}\eta _{ab} e^{2\sigma } \left[ {\left( {\partial _\alpha  a} \right) + \left\langle {\Xi }
 \mathrel{\left | {\vphantom {\Xi  {\partial _\alpha  \Xi }}}
 \right. \kern-\nulldelimiterspace}
 {{\partial _\alpha  \Xi }} \right\rangle } \right]\left[ {\left( {\partial ^\alpha  a} \right) + \left\langle {\Xi }
 \mathrel{\left | {\vphantom {\Xi  {\partial ^\alpha  \Xi }}}
 \right. \kern-\nulldelimiterspace}
 {{\partial ^\alpha  \Xi }} \right\rangle } \right]  \nonumber\\
 T_{\mu \nu }  &=& -\frac{1}{2}\left( {\partial _\mu  \sigma } \right)\left( {\partial _\nu  \sigma } \right) + \frac{1}{4}e^{2B\sigma } \delta _{\mu \nu } \left( {\partial _\alpha  \sigma } \right)\left( {\partial ^\alpha  \sigma } \right) \nonumber\\
 & & - G_{i\bar j} \left( {\partial _\mu  z^i } \right)\left( {\partial _\nu  z^{\bar j} } \right) + \frac{1}{2}e^{2B\sigma } \delta _{\mu \nu } G_{i\bar j} \left( {\partial _\alpha  z^i } \right)\left( {\partial ^\alpha  z^{\bar j} } \right) \nonumber\\
 & & + e^\sigma  \left\langle {\partial _\mu \Xi } \right|{\bf\Lambda} \left| {\partial _\nu \Xi } \right\rangle - \frac{1}{2}e^{\left(2B+1\right)\sigma } \delta _{\mu \nu }   \left\langle {\partial _\alpha \Xi } \right|{\bf\Lambda} \left| {\partial ^\alpha \Xi } \right\rangle \nonumber\\
  & &  - \frac{1}{2}e^{2\sigma } \left[ {\left( {\partial _\mu  a} \right) + \left\langle {\Xi }
 \mathrel{\left | {\vphantom {\Xi  {\partial _\mu  \Xi }}}
 \right. \kern-\nulldelimiterspace}
 {{\partial _\mu  \Xi }} \right\rangle } \right]\left[ {\left( {\partial _\nu  a} \right) + \left\langle {\Xi }
 \mathrel{\left | {\vphantom {\Xi  {\partial _\nu  \Xi }}}
 \right. \kern-\nulldelimiterspace}
 {{\partial _\nu  \Xi }} \right\rangle } \right]  \nonumber\\
  & & +  \frac{1}{4}e^{2\left(1+B\right)\sigma } \delta _{\mu \nu } \left[ {\left( {\partial _\alpha  a} \right) + \left\langle {\Xi }
 \mathrel{\left | {\vphantom {\Xi  {\partial _\alpha  \Xi }}}
 \right. \kern-\nulldelimiterspace}
 {{\partial _\alpha  \Xi }} \right\rangle } \right]\left[ {\left( {\partial ^\alpha  a} \right) + \left\langle {\Xi }
 \mathrel{\left | {\vphantom {\Xi  {\partial ^\alpha  \Xi }}}
 \right. \kern-\nulldelimiterspace}
 {{\partial ^\alpha  \Xi }} \right\rangle } \right].\label{StressTensor}
\eea

The universal axion's field equation (\ref{aEOM}) implies a solution of the form
\be
     da +\left\langle {\Xi }
 \mathrel{\left | {\vphantom {\Xi  {d\Xi }}}
 \right. \kern-\nulldelimiterspace}
 {{d\Xi }} \right\rangle = \alpha  e^{-2\sigma } dH,
\ee
where $H$ is an arbitrary function satisfying $\Delta H = 0$, and $\alpha  \in \mathbb{R}$. Similarly, the axions' field equation (\ref{AxionsEOM}) leads to
\be
    e^\sigma  \left| {{\bf\Lambda} d\Xi } \right\rangle  - \alpha dH\left| \Xi  \right\rangle  = \beta \left| {dK} \right\rangle \,\,\,\,\,{\rm where}\,\,\,\,\,\left| {\Delta K} \right\rangle  = 0\,\,\,\,\,{\rm and}\,\,\,\,\,\beta  \in \mathbb{R}.\label{Axion-K}
\ee

Since we are only interested in bosonic solutions, we consider the vanishing of the supersymmetric variations (\ref{SUSYHyperon1}, \ref{SUSYHyperon2}), which may be rewritten in matrix form as follows
\be
    \left[ {\begin{array}{*{20}c}
   {e^{\frac{\sigma }{2}} \left\langle {V}
    \mathrel{\left | {\vphantom {V {\partial _M  \Xi }}} \right. \kern-\nulldelimiterspace} {{\partial _M  \Xi }} \right\rangle  \Gamma ^M} & {} & {-\frac{1}{2}\left[ {\left( {\partial _M  \sigma } \right) - i \alpha e^{-\sigma }  {\left(\partial _M H\right) } } \right]\Gamma ^M}  \\
   {} & {} & {}  \\
   {\frac{1}{2}\left[ {\left( {\partial _N  \sigma } \right) + i \alpha e^{-\sigma }  {\left(\partial _N H\right) } } \right]\Gamma ^N} & {} & {e^{\frac{\sigma }{2}} \left\langle {{\bar V}}
    \mathrel{\left | {\vphantom {{\bar V} {\partial _N  \Xi }}} \right. \kern-\nulldelimiterspace} {{\partial _N  \Xi }} \right\rangle \Gamma ^N}  \\
\end{array}} \right]\left( {\begin{array}{*{20}c}
   {\epsilon _1 }  \\
   {}  \\
   {\epsilon _2 }  \\
\end{array}} \right) = 0
\ee
\be
    \left[ {\begin{array}{*{20}c}
   {e^{\frac{\sigma }{2}} e^{\hat ij} \left\langle {{U_j }}
    \mathrel{\left | {\vphantom {{U_j } {\partial _M  \Xi }}} \right. \kern-\nulldelimiterspace} {{\partial _M  \Xi }} \right\rangle \Gamma ^M } & {} & {-e_{\,\,\,\bar j}^{\hat i} \left( {\partial _M  z^{\bar j} } \right)\Gamma ^M}  \\
   {} & {} & {}  \\
   {e_{\,\,\,k}^{\hat j} \left( {\partial _N  z^k } \right)\Gamma ^N} & {} & {e^{\frac{\sigma }{2}} e^{\hat j\bar k} \left\langle {{U_{\bar k} }}
    \mathrel{\left | {\vphantom {{U_{\bar j} } {\partial _N  \Xi }}} \right. \kern-\nulldelimiterspace} {{\partial _M  \Xi }} \right\rangle \Gamma ^N}  \\
\end{array}} \right]\left( {\begin{array}{*{20}c}
   {\epsilon _1 }  \\
   {}  \\
   {\epsilon _2 }  \\
\end{array}} \right) = 0.
\ee

The vanishing of the determinants gives the BPS conditions:
\bea
 d\sigma  \wedge \star d\sigma  + \alpha ^2 e^{ - 2\sigma } dH \wedge \star dH + 4e^\sigma  \left\langle {V}
 \mathrel{\left | {\vphantom {V {d\Xi }}}
 \right. \kern-\nulldelimiterspace}
 {{d\Xi }} \right\rangle  \wedge \left\langle {{\bar V}}
 \mathrel{\left | {\vphantom {{\bar V} {\star d\Xi }}}
 \right. \kern-\nulldelimiterspace}
 {{\star d\Xi }} \right\rangle  &=& 0 \nonumber\\
 G_{i\bar j} dz^i  \wedge \star dz^{\bar j}  + e^\sigma  G^{i\bar j} \left\langle {{U_i }}
 \mathrel{\left | {\vphantom {{U_i } {d\Xi }}}
 \right. \kern-\nulldelimiterspace}
 {{d\Xi }} \right\rangle  \wedge \left\langle {{U_{\bar j} }}
 \mathrel{\left | {\vphantom {{U_{\bar j} } {\star d\Xi }}}
 \right. \kern-\nulldelimiterspace}
 {{\star d\Xi }} \right\rangle  &=& 0.\label{FromSYSY}
\eea

Using this with (\ref{DefOfRotInSympSpace}) we find
\be
    e^\sigma  \left\langle {d\Xi } \right|\mathop {\bf\Lambda} \limits_ \wedge  \left| {\star d\Xi } \right\rangle  = \frac{1}{2}d\sigma  \wedge \star d\sigma  + \frac{1}{2}\alpha ^2 e^{ - 2\sigma } dH \wedge \star dH + 2G_{i\bar j} dz^i  \wedge \star dz^{\bar j},\label{Rotation}
\ee
where we have used
\be
    \left\langle {d\Xi } \right.\mathop |\limits_ \wedge  \left. {\star d\Xi } \right\rangle  = 0,\label{ScalarProdOfXiStardXiEqualZero}
\ee
as required by the reality of the axions. The dilaton's equation (\ref{DilatonEOM}) then becomes
\be
    \left( {\Delta \sigma } \right)\star \mathbf{1} + \frac{1}{2}d\sigma  \wedge \star d\sigma  = \frac{1}{2}\alpha ^2 e^{ - 2\sigma } dH \wedge \star dH - 2G_{i\bar j} dz^i  \wedge \star dz^{\bar j}.\label{dilatonian}
\ee

Finally, the components of the Einstein equations reduce to
\bea
 \frac{1}{2}B\left( {3 - p} \right)\left[ {B\left( {2 - p} \right) - 1} \right]d\sigma  \wedge \star d\sigma  + \left[ {\frac{1}{2} - 2B\left( {3 - p} \right)} \right]G_{i\bar j} dz^i  \wedge \star dz^{\bar j}  \nonumber\\
  =  - \frac{1}{2}B\left( {3 - p} \right)\alpha ^2 e^{ - 2\sigma } dH \wedge \star dH \nonumber\\
 \frac{1}{2}B\left( {2 - p} \right)\left( {2B + 1} \right)d\sigma  \wedge \star d\sigma  + \left[ {2B\left( {2 - p} \right) - 1} \right]G_{i\bar j} dz^i  \wedge \star dz^{\bar j}  \nonumber\\
  = \frac{1}{2}B\left( {2 - p} \right)\alpha ^2 e^{ - 2\sigma } dH \wedge \star dH \nonumber\\
 \frac{1}{2}B\left( {2 - p} \right)\left[ {B\left( {1 - p} \right) - 1} \right]d\sigma  \wedge \star d\sigma  + \left[ {\frac{1}{2} - 2B\left( {2 - p} \right)} \right]G_{i\bar j} dz^i  \wedge \star dz^{\bar j}  \nonumber\\
  =  - \frac{1}{2}B\left( {2 - p} \right)\alpha ^2 e^{ - 2\sigma } dH \wedge \star dH.\label{EinsteinEquations}
\eea

It can be easily shown that the equations in (\ref{EinsteinEquations}) cannot be simultaneously satisfied for the case $\alpha=0$. They either lead to an imaginary $B$ or to trivial solutions with constant complex structure moduli \cite{Emam:2011uh}. On the other hand, the case of nonvanishing $\alpha$ leads to exactly two nontrivial solutions. These are $p=0,1$. In what follows, we study only the zero-brane case, deferring the study of the one-branes to future work.

\section{The Fields}\label{Fields}

For $p=0$, equations (\ref{Rotation}), (\ref{dilatonian}) and (\ref{EinsteinEquations}) are identically satisfied for any value of the constant $B$ if
\bea
 G_{i\bar j} dz^i  \wedge \star dz^{\bar j}  &=& 6B^2 d\sigma  \wedge \star d\sigma  \nonumber\\
 \left( {24B^2  - 4B + 1} \right)d\sigma  \wedge \star d\sigma  &=& \alpha ^2 e^{ - 2\sigma } dH \wedge \star dH \nonumber\\
 e^\sigma  \left\langle {d\Xi } \right|\mathop {\bf\Lambda} \limits_ \wedge  \left| {\star d\Xi } \right\rangle  &=& \left( {24B^2  - 2B + 1} \right)d\sigma  \wedge \star d\sigma  \nonumber\\
 \Delta e^{2B\sigma }  &=& 0.\label{p0solutionconditions}
\eea

The last equation of (\ref{p0solutionconditions}) implies the simple ansatz $e^{2B\sigma }  = H$, which leads to $B = {1 \mathord{\left/ {\vphantom {1 2}} \right. \kern-\nulldelimiterspace} 2}$ and $\alpha ^2  = 5$. Hence, the dilaton is fully specified in terms of $H$:
\be
 \sigma  = \ln H,
\ee
while the universal axion is, so far
\be
    da =  - \alpha dH^{ - 1}  - \left\langle {\Xi }
 \mathrel{\left | {\vphantom {\Xi  {d\Xi }}}
 \right. \kern-\nulldelimiterspace}
 {{d\Xi }} \right\rangle .
\ee

To find an expression for the axions, we look again at the vanishing of the hyperini transformations (\ref{SUSYHyperon1}) and (\ref{SUSYHyperon2}) and make the simplifying assumption $\epsilon_1=\pm\epsilon_2$. This leads to:
\bea
    \left\langle {V}
 \mathrel{\left | {\vphantom {V {d\Xi }}}
 \right. \kern-\nulldelimiterspace}
 {{d\Xi }} \right\rangle &=& \frac{1}{2}\left( {1 - i\alpha } \right)e^{ - \frac{\sigma }{2}} d\sigma\nonumber\\
    \left\langle {{\bar V}}
 \mathrel{\left | {\vphantom {{\bar V} {d\Xi }}}
 \right. \kern-\nulldelimiterspace}
 {{d\Xi }} \right\rangle &=& \frac{1}{2}\left( {1 + i\alpha } \right)e^{ - \frac{\sigma }{2}} d\sigma \nonumber\\
    \left\langle {{U_i }}
 \mathrel{\left | {\vphantom {{U_i } {d\Xi }}}
 \right. \kern-\nulldelimiterspace}
 {{d\Xi }} \right\rangle &=& e^{ - \frac{\sigma }{2}} G_{i\bar j} dz^{\bar j}\nonumber\\
    \left\langle {{U_{\bar j} }}
 \mathrel{\left | {\vphantom {{U_{\bar j} } {d\Xi }}}
 \right. \kern-\nulldelimiterspace}
 {{d\Xi }} \right\rangle &=& e^{ - \frac{\sigma }{2}} G_{i\bar j} dz^i.
\eea

These are the symplectic components of the full vector:
\bea
 \left| {d\Xi } \right\rangle  =& & \frac{1}{2}\left( {\alpha  - i} \right)e^{ - \frac{\sigma }{2}} d\sigma \left| V \right\rangle  + \frac{1}{2}\left( {\alpha  + i} \right)e^{ - \frac{\sigma }{2}} d\sigma \left| {\bar V} \right\rangle \nonumber \\
  & & + ie^{ - \frac{\sigma }{2}} dz^i \left| {U_i } \right\rangle  - ie^{ - \frac{\sigma }{2}} dz^{\bar j} \left| {U_{\bar j} } \right\rangle\nonumber\\
  =& & e^{ - \frac{\sigma }{2}} \mathfrak{Re} \left[ {\left( {\alpha  - i} \right)\left| V \right\rangle d\sigma  + 2i\left| {U_i } \right\rangle dz^i } \right].\label{dXi}
\eea

Clearly, the reality condition $\overline{\left| {d\Xi } \right\rangle}  = \left| {d\Xi } \right\rangle $ as well as the Bianchi identity on the axions are trivially satisfied. One can now substitute (\ref{dXi}) in (\ref{Axion-K}) to get
\bea
   \left| \Xi  \right\rangle d\sigma &=& \frac{1}{\alpha }  \left| {{\bf\Lambda} d\Xi } \right\rangle  - \frac{\beta }{\alpha }e^{-\sigma}\left| {dK} \right\rangle \nonumber\\
    &=& \frac{1}{\alpha }e^{ - \frac{\sigma }{2}} \mathfrak{Re}\left[\left(1+i\alpha\right)\left| V \right\rangle d\sigma\right] \nonumber\\
    &+& \frac{2}{\alpha }e^{ - \frac{\sigma }{2}} \mathfrak{Re}\left[\left| U_i \right\rangle dz^i\right]-\frac{\beta}{\alpha }e^{ - \sigma }\left| {dK} \right\rangle.\label{42}
\eea

The remaining field equations (\ref{ZZBarEOM}) are slightly simplified as a consequence of the third result of (\ref{p0solutionconditions}). They reduce to
\be
    \left( {\Delta z^i } \right)\star \mathbf{1} + \Gamma _{jk}^i dz^j  \wedge \star dz^k  = 0,\label{zEOMmodified}
\ee
and similarly for its complex conjugate counterpart. These cannot, however, be explicitly solved without knowledge of a metric on $\M_C$. However, one can conjecture several constructions. For instance, a direct dependence on $d\sigma$ can be assumed:
\be
    dz^i  = me^{A\sigma } f^i d\sigma,\label{dzAnsatz}
\ee
where $m$ and $A$ are arbitrary constants. Equation (\ref{zEOMmodified}) imposes the following constraint on the unknown functions $f^i$:
\be
    df^i  + me^{A\sigma } \Gamma _{jk}^i f^j f^k d\sigma  + \left( {A - 1} \right)f^i d\sigma  = 0\label{GeneralCY constraint}
\ee
which may further be simplified by the choice $m=A=1$ yielding the condition:
\be
    df^i  + \Gamma _{jk}^i f^j f^k de^\sigma   = 0.\label{DiffCon}
\ee

Adopting this ansatz, equation (\ref{42}) can now be rewritten as
\be
   \left| \Xi  \right\rangle
    = \frac{1}{\alpha }e^{ - \frac{\sigma }{2}} \mathfrak{Re}\left[\left(1+i\alpha\right)\left| V \right\rangle \right] \nonumber\\
    + \frac{2}{\alpha }e^{  \frac{\sigma }{2}} \mathfrak{Re}\left[f^i\left| U_i \right\rangle \right],\label{42again}
\ee
where, without loss of generality, we have chosen $\beta=0$. The first equation in (\ref{p0solutionconditions}) leads to the additional constraint:
\be
    G_{i\bar j} f^i f^{\bar j}  = \frac{3}{{2m^2 }}e^{ - 2A\sigma }  = \frac{3}{2}e^{ - 2\sigma }.
\ee

Using these results, we find
\bea
 \left\langle {\Xi }
 \mathrel{\left | {\vphantom {\Xi  {d\Xi }}}
 \right. \kern-\nulldelimiterspace}
 {{d\Xi }} \right\rangle  &=&  -  {\frac{{7}}{{2\alpha }}} dH^{ - 1} \,\,\,\,{\rm leading}\,\,{\rm to} \nonumber\\
 da &=& \left( {\frac{{7 - 2\alpha ^2 }}{{2\alpha }}} \right)dH^{ - 1} \,\,\,\,\,{\rm and} \nonumber\\
 a &=& c - \left(\frac{2\alpha^2-7}{2\alpha}\right)\frac{1}{H},
\eea
where $c$ is an arbitrary integration constant related to the asymptotic value of $a$. Finally, solving the SUSY condition $\delta \psi = 0$ gives the following form for the spinors
\bea
    \epsilon _1  &=& e^{n\sigma  + \Upsilon } \hat \epsilon, \,\,\,\,{\rm where}\nonumber\\
 n &=& \frac{1}{2}\left[ { \pm 1 + i\alpha \left( { \pm 1 - \frac{1}{2}} \right)} \right] \nonumber\\
 \left( {\partial _\mu  \Upsilon } \right) &=& \frac{1}{4}\left[ {Y_\mu \pm \left( {\partial _\nu  \sigma } \right)\varepsilon _\mu ^{\,\,\,\,\,\nu } } \right],
\eea
and $\hat \epsilon$ is a constant spinor. We have used
\be
    \omega _\alpha ^{\;\;\hat \beta \hat \gamma }= \frac{1}{2}\left( {\delta _\alpha ^{\hat \beta }
    \delta ^{\hat \gamma \rho }  - \delta ^{\hat \beta \rho } \delta _\alpha ^{\hat \gamma
    }     } \right)\left( {\partial _\rho  \sigma } \right), \quad\quad
    D_\mu   = \partial _\mu   + \frac{1}{4}\left( {\partial _\nu  \sigma
    }     \right){\Gamma _\mu } ^\nu,
\ee
as well as the Dirac matrices projection conditions\footnote{The Einstein summation convention is \emph{not} used over the index
$s$.}:
\begin{eqnarray}
    \Gamma _{\hat \mu \hat \nu } \epsilon _s  &= b_s \varepsilon _{\hat
    \mu     \hat \nu } \epsilon _s,  \quad\quad\quad\quad s &=(1,2),\quad b_s=\pm
    c \nonumber \\
    \Gamma _\mu^{\;\;\;\nu} \epsilon _s  &= b_s {\varepsilon_\mu}^{\;\nu}\epsilon
    _s, \quad\quad
    \Gamma ^\mu  \epsilon _s  &=  - b_s {\varepsilon_\nu}^{\;\mu} \Gamma
    ^\nu          \epsilon _s.\label{projection}
\end{eqnarray}

If we now solve the Laplace equation $\Delta H=0$ to find
\be
    H\left( r\right) = 1 + \frac{q}{{r^2 }}\quad\quad\quad {\rm where}   \quad\quad\quad q\in \mathbb{R}
\ee
and $r$ is the usual radial coordinate in 4-D space, then the zero-brane coupled to the hypermultiplets can be represented by:
\bea
    ds^2  &=&  - dt^2  + \left( {1 + \frac{q}{{r^2 }}} \right)\left( {dr^2  + r^2 d\Omega ^2_3} \right)     \nonumber\\
    \sigma \left( r \right) &=& \ln \left( {1 + \frac{q}{{r^2 }}} \right)\nonumber\\
    a &=& a_\infty   \pm \frac{{3q}}{{2\sqrt 5 \left( {r^2  + q} \right)}}\nonumber\\
    dz^i  &=& -2q f^i\frac{{dr}}{{r^3 }}\,\,\,\,\,\,{\rm such}\,\,{\rm that}\,\,\,\,\,\,df^i  - 2q\Gamma _{jk}^i f^j f^k\frac{dr}{r^3} =0\,\,\,\,\,\,{\rm and}\,\,\,\,\,\,G_{i\bar j} f^i f^{\bar j}  = \frac{{3r^4 }}{{2\left( {r^2  + q} \right)^2 }}    \nonumber\\
    \left| \Xi  \right\rangle  &=&   \frac{{r }}{{\sqrt {5\left( {r^2  + q} \right)} }}\mathfrak{Re}\left[ {\left( {1 \pm i\sqrt 5 } \right)\left| V \right\rangle } \right] + \frac{2}{{r }}\sqrt {\frac{{\left( {r^2  + q} \right)}}{5}} \mathfrak{Re}\left[ {f^i \left| {U_i } \right\rangle } \right]\nonumber\\
    \left| {d\Xi } \right\rangle  &=& \pm 2q\mathfrak{Re} \left[ {\frac{{\left( { \pm \sqrt 5  - i} \right)}}{{\left( {r^2  + q} \right)^{\frac{3}{2}} }}\left| V \right\rangle  + \frac{{2i}}{{r^2\sqrt {r^2  + q} }}f^i \left| {U_i } \right\rangle } \right]dr,\label{FullSolution}
\eea
where $d\Omega ^2_3$ is the unit $S^3$ metric. The equations in (\ref{FullSolution}) represent a full symplectic solution, but only a partial spacetime one. The entire construction is based on the choice that the dilaton and the universal axion are independent of the moduli, and that the entire moduli dependence is carried exclusively by the axions, while the moduli themselves are dependent on an unknown symplectic scalar $f^i$. The condition $df^i  - 2q\Gamma _{jk}^i f^j f^k\frac{dr}{r^3} =0$ is interesting. While there are no guarantees that there exists a CY submanifold that satisfies it, or even the more general (\ref{GeneralCY constraint}), we may argue that it is in fact a compact version of the more complicated attractor equations found in other solutions and as such is at least as possible to satisfy as they are.

The quantity $q$ is a coupling constant relating the behavior of the fields to each other and to gravity. Since the metric is asymptotically flat; the ADM mass of the brane is easily calculable and is clearly proportional to $q$. Since the value of $q$ can be either positive or negative, we note the following: For positive values of $q$ the solution is entirely smooth between the central singularity and infinity. While for the case of negative $q$, a curvature singularity exists at $r=\sqrt{\left| q \right|}$. As such the negative $q$ result has two singularities, one at $r=0$ and the other constituting an $S^3$ surface with radius $r=\sqrt{\left| q \right|}$. In both cases the singularities are naked; no horizons exist.

\section{Conclusion}

The primary objective of this work was to apply the methods developed in our earlier paper \cite{Emam:2009xj} and construct $D=5$ hypermultiplet fields in a specific spacetime background, simply by exploiting the symplectic symmetry of the theory and finding solutions that are based on symplectic invariants and vectors. In so doing, we have also shown that only two (Poincar\'{e})$_{p+1} \times SO\left(4-p\right)$ backgrounds are allowed (within the symmetries assumed). Focusing on one of these possibilities, we constructed a zero-brane coupled to the hypermultiplet fields of $\N=2$ supergravity. The metric and fields are well behaved in the far field region and are dependent on the ADM mass of the brane. We found explicit expressions for the metric, dilaton and the universal axion. On the other hand the axions are dependent on spacetime-unspecified symplectic basis vectors and the moduli are proportional to an unknown set of functions $f^i$, satisfying specific conditions, which we also derived. What we have then is a complete symplectic solution, but a partial spacetime one. Clearly, a full solution hinges on the values of $f^i$, i.e. on solving the aforementioned constraints, or equivalently on solving (\ref{zEOMmodified}). This is unlikely to be possible without a full understanding of the structure of the CY submanifold. In reverse, however, further study of these functions may provide clues to the underlying manifold. Although we have focused on the $p=0$ solution, the setup investigated here also admits a $p=1$ configuration. We plan to continue in this direction and study the possible constraints on the moduli (similar to $f^i$) that should arise.

\pagebreak

\end{document}